\documentclass[twocolumn,prb,showpacs]{revtex4}
\usepackage{graphicx}
\usepackage{dcolumn}
\usepackage{bm}
\usepackage{subfigure}
\usepackage{amsmath}
\usepackage{hhline}
\begin{document}
\title{Resonant backward scattering of light by a two-side-open subwavelength metallic slit}
\author{S. V. Kukhlevsky$^a$, M. Mechler$^b$, L. Csap\'o$^c$, K. Janssens$^d$, O. Samek$^e$}
\affiliation{$^a$Institute of Physics, University of P\'ecs, Ifj\'us\'ag u.\ 6, P\'ecs 7624, Hungary\\
$^b$South-Trans-Danubian Cooperative Research Centre, University of P\'ecs, Ifj\'us\'ag u.\ 6, P\'ecs 7624, Hungary\\
$^c$Institute of Mathematics and Information, University of P\'ecs, Ifj\'us\'ag u.\ 6, P\'ecs 7624, Hungary\\
$^d$Department of Chemistry, University of Antwerp, Universiteitsplein 1, B-2610 Antwerp, Belgium\\
$^e$Institute of Spectrochemistry and Applied Spectroscopy, Bunsen-Kirchhoff-Str.\ 11, D-44139 Dortmund, Germany}
\begin{abstract}
The backward scattering of TM-polarized light by a two-side-open subwavelength slit in a metal
film is analyzed. We show that the reflection coefficient versus wavelength possesses a
Fabry-Perot-like dependence that is similar to the anomalous behavior of transmission reported
in the study [Y. Takakura, Phys. Rev. Lett. \textbf{86}, 5601 (2001)]. The open slit totally
reflects the light at the near-to-resonance wavelengths. In addition, we show that the
interference of incident and resonantly backward-scattered light produces in the near-field
diffraction zone a spatially localized wave whose intensity is 10-10$^3$ times greater than
the incident wave, but one order of magnitude smaller than the intra-cavity intensity.
The amplitude and phase of the resonant wave at the slit entrance and exit are different
from that of a Fabry-Perot cavity.
\end{abstract}
\maketitle
\section{Introduction}
The most impressive features of light scattering by subwavelength
metallic nanostructures are resonant enhancement and localization
of the light by excitation of electron waves in the metal (for
example, see
refs.~\cite{Neer,Harr,Betz1,Ebbe,Hess,Nev,Nev1,Sarr1,Cscher,Port,Trea,Asti,Pop,Bozs,Taka,Yang,Hibb,Gar3,Cao,
Barb,Dykh,Stee,Shi,Scho,Naha,Bouh,Kuk2,Gar1,Lind,Xie,Decha,Bori,Fan,Zay,Li,Lab,Ben,Monz,Vigo}).
In the last few years, a great number of studies have been devoted
to the nanostructures in metal films, namely a single aperture, a
grating of apertures and an aperture surrounded by grooves.
Since the recent paper of Ebbesen and colleagues\cite{Ebbe} on the
resonantly enhanced transmission of light observed for a 2D array
of subwavelength holes in metal films, the resonant phenomenon is
intensively discussed in the
literature.\cite{Ebbe,Hess,Nev,Nev1,Sarr1,Cscher,Port,Trea,Asti,Li,Pop,Bozs,Taka,Yang,Hibb,Gar3,Cao,
Barb,Dykh,Stee,Shi,Scho,Naha,Bouh,Kuk2,Gar1,Lind,Xie,Decha,Bori,Fan,Zay,Monz,Vigo}
Such a kind of light scattering is usually called a Wood's
anomaly. In the early researches, Hessel and Oliner showed that
the resonances come from coupling between nonhomogeneous
diffraction orders and eigenmodes of the grating.\cite{Hess}
Neviere and co-workers discovered two other possible origins of
the resonances.\cite{Nev,Nev1} One appears when the surface
plasmons of a metallic grating are excited. The other occurs when
a metallic grating is covered by a dielectric layer, and
corresponds to guided modes resonances in the dielectric film. The
role of resonant Wood's anomalies and Fano's profiles in the
resonant transmission were explained in the study.\cite{Sarr1}

The phenomena involved in propagation through hole arrays are
different from those connected with slit arrays. In a slit
waveguide there is always a propagating mode inside the channel,
while in a hole waveguide all modes are evanescent for hole
diameters smaller than approximately a wavelength. In the case of
slit apertures in a thick metal film, the transmission exhibits
enhancement due to a pure geometrical reason, the resonant
excitation of propagating modes inside the slit
waveguide.\cite{Port,Hibb,Taka,Asti,Cao} At the resonant
wavelengths, the transmitted field increases via the strong
coupling of an incident wave with the waveguide modes giving a
Fabry-Perot-like behavior.\cite{Asti,Taka,Yang,Kuk2} In the case
of films, whose thickness are too small to support the
intra-cavity resonance, the extraordinary transmission can be
caused by another mechanism, the generation of resonant surface
plasmon polaritons and coupling of them into
radiation.\cite{Ebbe,Port,Hibb,Gar1,Cscher} Both physical
mechanisms play important roles in the extraordinary transmission
through arrays of two-side-open slits (transmission gratings) and
the resonant reflection by arrays of one-side-open slits
(reflection gratings). A model of trapped (waveguide) modes has
been recently used to show that an array of two-side-open slits
can operate like a reflection grating totally reflecting
TE-polarized light.\cite{Bori} The surface plasmons and Rayleigh
anomalies were involved in explanation of reflective properties of
such a kind of gratings.\cite{Stee}

The studies \cite{Taka,Yang,Gar3,Kuk2} have pointed out that the
origin of anomalous scattering of light by a grating of slits
(holes) can be better understood by clarifying the transmission
and reflection properties of a single subwavelength slit. Along
this direction, it was already demonstrated that the intensity of
TM-polarized light resonantly transmitted through a single slit
can be 10-10$^3$ times higher than the incident
wave\cite{Neer,Harr,Betz1,Kuk2} and that the transmission
coefficient versus wavelength possesses a Fabry-Perot-like
behavior\cite{Taka,Yang,Kuk2}. Unfortunately, the reflection
properties of the slit have received no attention in the
literature. The very recent study\cite{Bori} only concerned the
problem by regarding the total reflection of TE-polarized light by
a grating of two-side-open slits to properties of the independent
slit emitters.

In this article, the backward scattering of light by a
two-side-open subwavelength slit is analyzed. To compare
properties of the light reflection with the extraordinary
transmission\cite{Taka,Yang,Kuk2}, we consider the scattering of
TM-polarized light by a slit in a thick metallic film of perfect
conductivity. From the latter metal property it follows that
surface plasmons do not exist in the film. Such a metal can be
described by the Drude model for which the plasmon frequency tends
towards infinity. The traditional approach based on the Neerhoff
and Mur solution of Maxwell's equations is used in the
computations.\cite{Neer,Harr,Betz1} The article is organized as
follows. The theoretical background, numerical analysis and
discussion are presented in Section~II. The summary and
conclusions are given in Section~III. The brief description of the
model is presented in the Appendix.

\section{Numerical analysis and discussion}
\begin{figure}[tb]
\begin{center}
\includegraphics[keepaspectratio,width=\columnwidth]{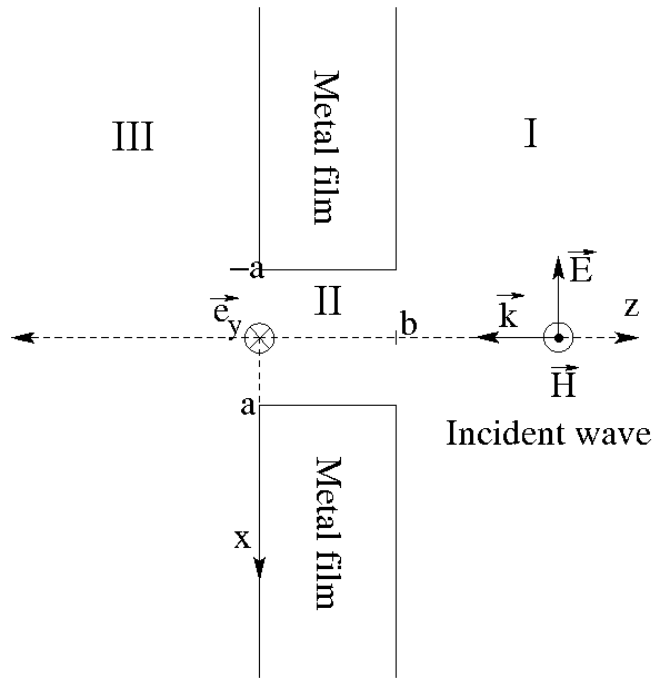}
\end{center}
\caption{\label{fig:5}Propagation of a continuous wave through a
subwavelength nano-sized slit in a thick metal film.}
\end{figure}

It is well known that when a light wave is scattered by a
subwavelength metallic object, a significant part of the incident
light can be scattered backward (reflected) whatever the object be
reflecting or transparent. It was recently demonstrated that an
array of two-side-open subwavelength metallic slits effectively
reflects light waves at the appropriate resonant
conditions.\cite{Stee,Bori} One may suppose that this is true also
in the case of a single slit. In this section, we test whether a
light wave can be resonantly reflected by a single two-side-open
subwavelength metallic slit. To address this question, the energy
flux in front of the slit is analyzed numerically for various
regimes of the light scattering. In order to compare properties of
the light reflection with that of the extraordinary (resonant)
transmission\cite{Taka,Yang,Kuk2}, we consider the zeroth-order
scattering of a time-harmonic wave of TM-polarized light by a slit
in a perfectly conducting thick metal film placed in vacuum (Fig.
\ref{fig:5}).
The energy flux $\vec S_I$ in front of the slit is compared with
the fluxes $\vec S_{II}$ and $\vec S_{III}$ inside the slit and
behind the slit, respectively. The amplitude and phase of the
light wave at the slit entrance and exit are compared with that of
a Fabry-Perot cavity. The electric $\vec{E}$ and magnetic
$\vec{H}$ fields of the light are computed by using the
traditional approach based on the Neerhoff and Mur solution of
Maxwell's equations.\cite{Neer,Harr,Betz1} For more details of the
model, see the Appendix.

According to the model, the electric $\vec E(x,z)$ and magnetic
$\vec H(x,z)$ fields in front of the slit (region~I), inside the
slit (region~II) and behind the slit (region~III) are determined
by the scalar fields $U_1(x,z)$, $U_2(x,z)$ and $U_3(x,z)$,
respectively. The scalar fields are found by solving the Neerhoff
and Mur integral equations. The magnetic field of the wave is
assumed to be time harmonic and constant in the $y$ direction:
$\vec{H}(x,y,z,t)=U(x,z)\exp(-i\omega{t})\vec{e}_y$. In front of
the slit, the field is decomposed into
$U_1(x,z)=U^i(x,z)+U^r(x,z)+U^d(x,z)$. The field $U^i(x,z)$
represents the incident field, which is assumed to be a plane wave
of unit amplitude; $U^r(x,z)$ denotes the field that would be
reflected if there were no slit in the film; $U^d(x,z)$ describes
the backward diffracted field due to the presence of the slit. The
time averaged Poynting vector (energy flux) $\vec S$ of the
electromagnetic field is calculated (in CGS units) as
$\vec{S}=(c/16\pi)(\vec{E}\times\vec{H}^*+\vec{E}^*\times\vec{H})$.
The reflection coefficient $R=S_{int}^{rd}$ is given by the
normalized flux $S_n^{rd}=S^{rd}/S^i$ integrated over the slit
width $2a$ at the slit entrance ($z=b$), where $S^{rd}$ is the $z$
component of the backward scattered flux, and $S^i$ is the
incident flux along the $z$ direction. The flux
$S^{rd}=S^{rd}(U^r,U^d)$ is produced by the interference of the
backward scattered fields $U^r(x,z)$ and $U^d(x,z)$. The
transmission coefficient $T=S_{int}^3(b)$ is determined by the
normalized flux $S_n^3=S^3/S^i$ integrated over the slit width at
the slit exit ($z=0$), where the flux $z$-component $S^3 =
S^3(U^3)$ is produced by the forward scattered (transmitted) field
$U^3(x,z)$. Notice that the definitions of the reflection $R$ and
transmission $T$ coefficients are equivalent to the more
convenient ones defined as the integrated reflected or transmitted
flux divided by the integrated incident flux. In the following
analysis, the reflection and transmission coefficients are
compared to the fluxes $S_{int}^d$ and $S_{int}^{ird}$ obtained by
integrating the normalized fluxes $S_n^d=S^d/S^i$ and
$S_n^{ird}=S^{ird}/S^i$, respectively.

\begin{figure}[!tb]
\includegraphics[keepaspectratio,width=\columnwidth]{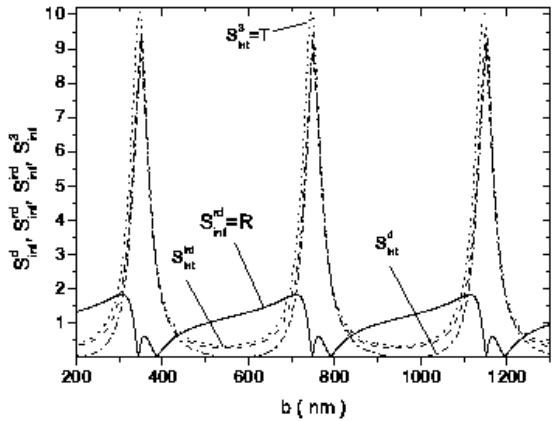}
\subfigure{\label{fig:1a}(a)}
\includegraphics[keepaspectratio,width=\columnwidth]{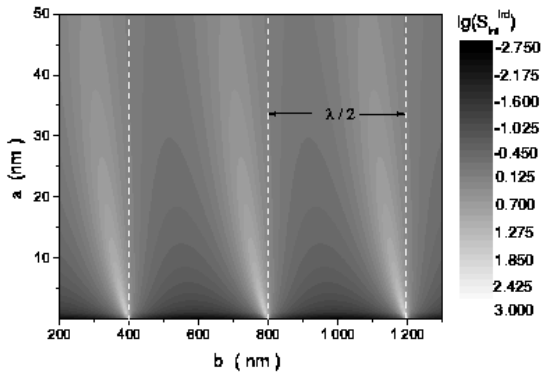}
\subfigure{\label{fig:1b}(b)}
\caption{
(a) The reflection coefficient $R=S_{int}^{rd}(b)$, the
transmission coefficient $T=S_{int}^3(b)$, and the integrated
fluxes $S_{int}^d(b)$ and $S_{int}^{ird}(b)$ as a function of the
film thickness $b$ computed for the wavelength $\lambda=800$~nm
and the slit width $2a=25$~nm. (b) The logarithm of the integrated
flux $S_{int}^{ird}(a,b)$ as a function of the slit half-width $a$
and film thickness $b$.}
\end{figure}

We analyzed the backward scattering of light for a wide range of
scattering conditions determined by values of the wavelength
$\lambda$, slit width $2a$ and film thickness $b$. As an example,
the reflection coefficient $R=S_{int}^{rd}(b)$ as a function of
the film thickness $b$ computed for the wavelength
$\lambda=800$~nm and the slit width $2a=25$~nm is shown in
Fig.~\ref{fig:1a}.
The transmission coefficient $T=S_{int}^3(b)$ and the integrated
fluxes $S_{int}^d(b)$ and $S_{int}^{ird}(b)$ are presented in the
figure for the comparison. We note the reflection resonances of
$\lambda/2$ periodicity with the maxima $R_{max}\approx{2}$. In
agreement with the previous results\cite{Taka,Yang,Kuk2}, one can
see also the transmission resonances having the same period, and
the peak heights $T\approx{10}$ ($T\approx\lambda/2\pi a$) at the
resonances. It is worth to note the correlation between the
positions of maxima and minima in the reflection and transmission.
The resonance positions for the total reflection are somewhat
left-shifted with respect to the transmission resonances. The
maxima of the transmission coefficient correspond to reflection
minima. In Fig.~\ref{fig:1a}, one can observe also many satellite
peaks in reflection. For one broad minimum, it appears a local
reflection maximum, which is characterized by a weak amplitude.
The local maxima appear before 400, 800, and 1200~nm. The
positions of the local maxima approximately correspond to the
$S^d_{int}$ maxima. To clarify a role of the fields $U^i$, $U^r$,
$U^d$ and $U^3$ in the resonant backward scattering, we compared
the integrated flux $S_{int}^{ird}(U^i,U^r,U^d)$ with the fluxes
$S_{int}^d(U^d)$ and $S_{int}^3(U^3)=T$. One can see from
Fig.~\ref{fig:1a} that the flux $S_{int}^{ird}$ produced in front
of the slit by the interference of the incident field $U^i(x,z)$
and the backward scattered fields $U^r(x,z)$ and $U^d(x,z)$ is
practically undistinguishable from that generated by the backward
diffracted field $U^d$ and forward scattered (transmitted) field
$U^3$. The integrated flux $S_{int}^{ird}(a,b)$ as a function of
the slit half-width~$a$ and film thickness $b$ is shown in
Fig.~\ref{fig:1b}. We notice that the widths and shifts of the
resonances increase with increasing the value $a$. Analysis of
Fig.~\ref{fig:1a} indicates that the difference between the
integrated fluxes $S_{int}^{rd}(U^r,U^d)=R$ and $S_{int}^3(U^3)=T$
($T{\approx}S_{int}^d(U^d)$) appears due to the interference of
the backward diffracted field $U^d(x,z)$ and the reflected field
$U^r(x,z)$.

\begin{figure}[tb]
\includegraphics[keepaspectratio,width=\columnwidth]{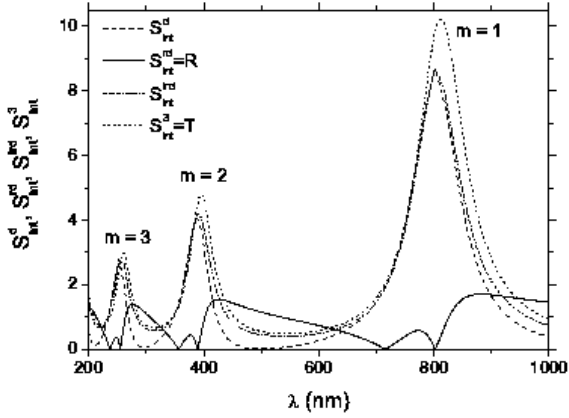}
\subfigure{\label{fig:2a}(a)}
\includegraphics[keepaspectratio,width=\columnwidth]{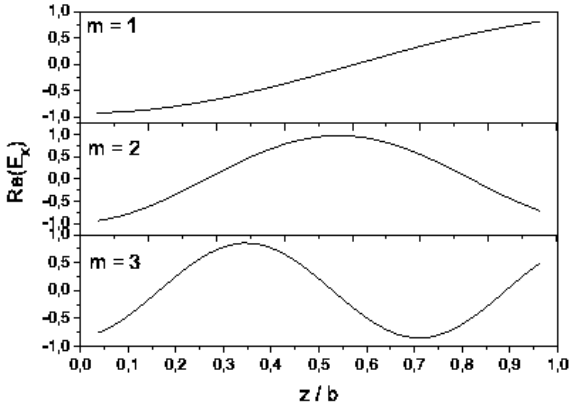}
\subfigure{\label{fig:2b}(b)}
\caption{
(a) The reflection coefficient $R=S_{int}^{rd}(\lambda)$,
the transmission coefficient $T=S_{int}^3(\lambda)$, and the integrated
fluxes $S_{int}^d(\lambda)$ and $S_{int}^{ird}(\lambda)$ versus the
wavelength $\lambda$ computed for the slit width $2a=25$~nm and the
screen thickness $b=351$~nm.
(b) The real part of the normalized electric field x-component
$E_x(U_2)=E_x(x,z)$ versus the normalized distance $z/b$ inside the
slit cavity at $x=0$, for the resonant wavelength $\lambda_r^1=800$~nm, $\lambda_r^2=389$~nm, $\lambda_r^3=255$~nm.}
\end{figure}

The dispersion of the reflection coefficient
$R(\lambda)=S_{int}^{rd}(\lambda)$ for the slit width $2a=25$~nm
and the screen thickness $b=351$~nm is shown in Fig.~\ref{fig:2a}.
The integrated fluxes $S_{int}^d(\lambda)$,
$S_{int}^{ird}(\lambda)$ and $S_{int}^3(\lambda)=T(\lambda)$
versus the wavelength are shown in the figure for the comparison.
A very interesting behavior of the dispersion is that the
coefficient $R$ versus the wavelength $\lambda$ possesses a
Fabry-Perot-like dependence that is similar to the anomalous
behavior of transmission $T(\lambda)$ reported in the
studies\cite{Asti,Taka,Yang,Gar3,Kuk2}. In agreement with the
studies\cite{Neer,Harr,Betz1,Kuk2}, the height of the first
(maximum) transmission peak is given by $T\approx \lambda_r^1/2\pi
a$. The wavelengths corresponding to the resonant peaks
$\lambda_r^m\approx 2b/m$ ($m=1,2,3,\dots$) are in accordance with
the results\cite{Taka}. The high peak amplitudes (enhancement),
however, are different from the low magnitudes (attenuation)
predicted in the study\cite{Taka}, but compare well with the
experimental and theoretical
results\cite{Neer,Harr,Betz1,Kuk2,Yang}. The difference is caused
by the manner in which the Maxwell equations are solved. The
study\cite{Taka} uses a simplified approach based on the matching
the cavity modes expansion of the light wave inside the slit with
the plain waves expansion above and below the slit using two
boundary conditions, at $z=0$ and $|z|=b$. Conversely, the
Neerhoff and Mur method performs the matching with five boundary
conditions, at $z\rightarrow 0$, $z\rightarrow b$, $x\rightarrow
a$, $x\rightarrow -a$, and $r\rightarrow \infty$. In contrast to
the sharp Lorentzian-like transmission peaks, the slit forms very
wide Fano-type reflection bands (see, Fig.~\ref{fig:2a}). For one
broad minimum in reflection, it appears also a local reflection
maximum, which is characterized by weak amplitude. At the
near-to-resonance wavelengths of the transmission, the open
aperture totally reflects the light. It is worth to note the
correlation between the wavelengths for maxima and minima in the
reflection $R(\lambda)$, transmission $T(\lambda)$, and the flux
$S^d_{int}$ (Table~\ref{tab:refl}).
\begin{table}[h]
\begin{center}
\begin{tabular}{cccc}
\hhline{====} $\lambda$ (nm) of&$R_{max}^{main}$
&$R_{max}^{little}$&$R_{min}$\\\hline
&276&248&237\\
&426&377&255\\
&882&773&356\\
&&&389\\
&&&714\\
&&&802\\
\hhline{====} $\lambda$ (nm) of&$T_{max}$&$T_{min}$
&$S^d_{max}$\\\hline
&260&226&253\\
&396&315&388\\
&812&542&802\\\hhline{====}
\end{tabular}
\end{center}
\caption{\label{tab:refl}The wavelengths for maxima and minima in
the reflection $R(\lambda)$, transmission $T(\lambda)$, and the
flux $S^d_{int}$.}
\end{table}
The resonance wavelengths for the main reflection maxima are
red-shifted with respect to the transmission resonances. The
wavelengths of both the transmission and reflection (main and
little) resonances are red-shifted with respect to the Fabry-Perot
wavelengths $\lambda_r^m=702$~nm, $351$~nm,$\dots$
($\lambda_r^m\approx 2b/m$, $m=1,2,3,\dots$). To understand the
physical mechanism of the resonant backward scattering, we also
compared the integrated flux $S_{int}^{ird}(U^i,U^r,U^d)$ with the
fluxes $S_{int}^d(U^d)$ and $S_{int}^3(U^3)=T$. As can be seen
from Fig.~\ref{fig:2a}, the integrated fluxes
$S_{int}^d(\lambda)$, $S_{int}^{ird}(\lambda)$ and
$S_{int}^3(\lambda)=T(\lambda)$ are practically undistinguishable
also in the $\lambda$-domain (for the $b$-domain, see
Fig.~\ref{fig:1a}). The difference between the integrated fluxes
$S_{int}^{rd}(U^r,U^d)=R$ and $S_{int}^3(U^3)=T$
($T{\approx}S_{int}^d(U^d)$) is caused by the interference of the
backward diffracted field $U^d(x,z)$ and the reflected field
$U^r(x,z)$ in the energy flux
$\vec{S}\sim{(\vec{E}\times\vec{H}^*+\vec{E}^*\times\vec{H})}$.
The wavelengths of the little maxima of the reflection
$R=R(U^d,U^r)$ correspond approximately to the high maxima
$S^d_{int}$. Therefore, the little maxima can be attributed to the
interference of the reflected field $U^r$ with the dominant
diffracted field $U^d$. The red shifts and the asymmetrical shapes
of the reflection bands can be explained by a Fano
analysis\cite{Sarr1} of the scattering problem by distinguishing
resonant and non-resonant interfering contributions to the
reflection process. The resonant contribution is given by the
field $U^d$ and the non-resonant one is attributed to the field
$U^r$. Other interesting interpretations of the shifts of resonant
wavelengths in the transmission spectra from the values $2b/m$ can
be found in the studies\cite{Asti,Taka,Yang,Kuk2,Bori,Trea}. It
should be mentioned that the asymmetrical behavior of reflection
was observed also in the case of a Fabry-Perot
resonator\cite{Monz,Vigo}. The conditions to achieve such an
asymmetry are rested on the existence of dissipative loss in the
resonator. There is no explicit loss in the present problem, but
the dissipative loss can be substituted by radiative loss due to
the diffraction by the slit.

After the analysis of Fig.~\ref{fig:1a}, it is not surprising that
the maxima of the transmission are accompanied by the minima of
the reflection also in the $\lambda$-domain (see,
Fig.~\ref{fig:2a}). It should be noted in this connection that
such a behavior of $R(\lambda)$ and $T(\lambda)$ is similar to
that observed in the case of excitation of the surface plasmons in
an array of slit in a thin metal film.\cite{Stee} In the
study\cite{Stee}, the minima in reflection spectra corresponding
to the maxima in the transmission spectra were attributed to the
redistribution of the energy of diffracted evanescent order into
the propagating order. In the case of a thick film, we explain
such a behavior by another physical mechanism, the interference of
the backward diffracted field $U^d(x,z)$ and the reflected field
$U^r(x,z)$. It can be noted that the correlation of positions of
reflection minima and transmission maxima (see, Figs.~\ref{fig:1a}
and \ref{fig:2a}) are consistent with that predicted by the
study\cite{Bori} for TE-polarized light scattered by a grating of
two-side-open slits in a thick metal film. However, the values of
$R(\lambda)$ and $T(\lambda)$ are in contrast to the relation
$R(\lambda)+T(\lambda)=1$ given in the study\cite{Bori}. The
difference can be explained by the fact that we examined light
scattering by an infinite screen using local definitions of $R$
and $T$, while the study\cite{Bori} analyzed the global reflection
and transmission by a grating of finite size.

\begin{figure}[tb]
\includegraphics[keepaspectratio,width=\columnwidth]{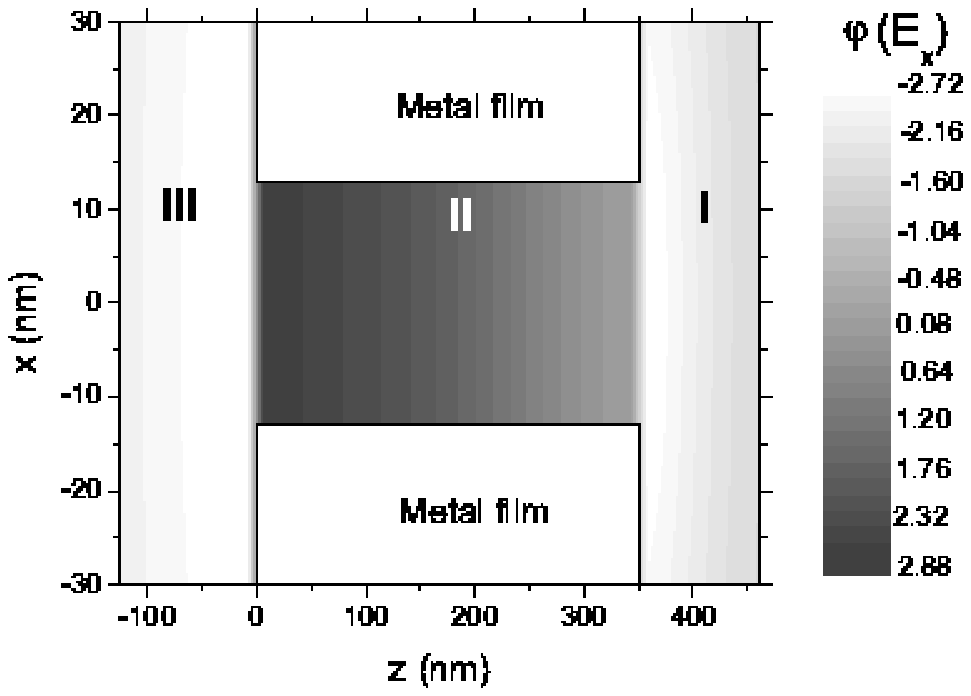}
\subfigure{\label{fig:3a}(a)}
\includegraphics[keepaspectratio,width=\columnwidth]{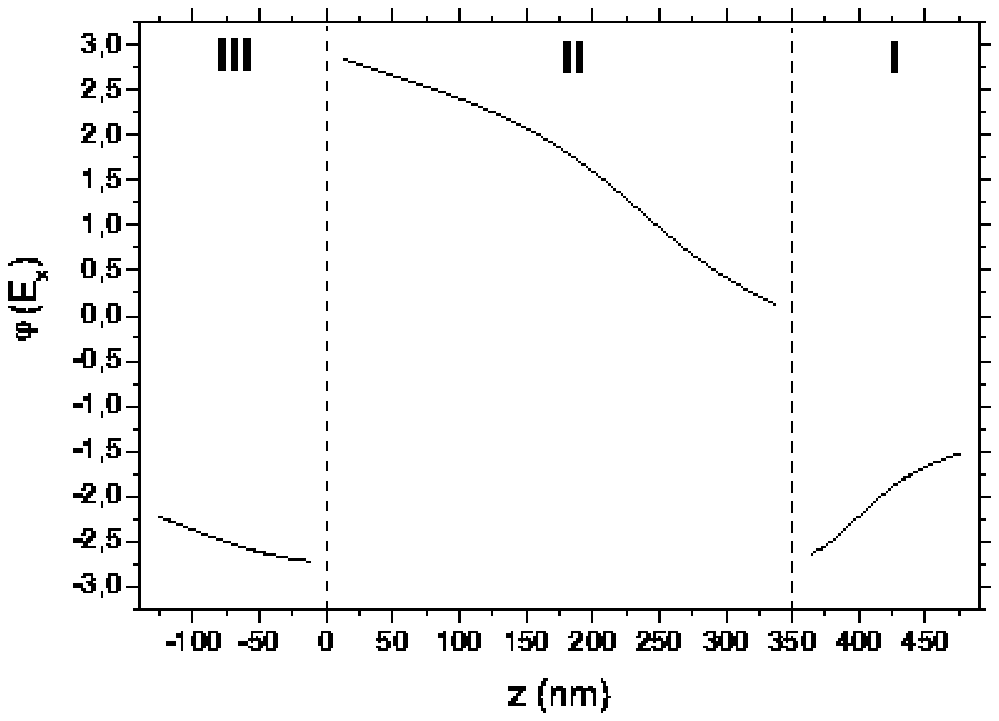}
\subfigure{\label{fig:3b}(b)} \caption{\label{fig:3}(a) The phase
distribution $\varphi(x,z)$ of the electric field $x$-component
$E_x(x,z)$ inside and outside the slit. The field component
$E_x(x,z)$ is given by $E_x(U^i,U^r,U^d)$, $E_x(U^2)$ and
$E_x(U^3)$ in the regions I, II and III, respectively. (b) The
phase distribution $\varphi(x,z)$ at $x=0$. The slit width
$2a=25$~nm, the film thickness $b=351$~nm and the wavelengths
$\lambda=800$~nm.}
\end{figure}
\begin{figure}[tb]
\includegraphics[keepaspectratio,width=\columnwidth]{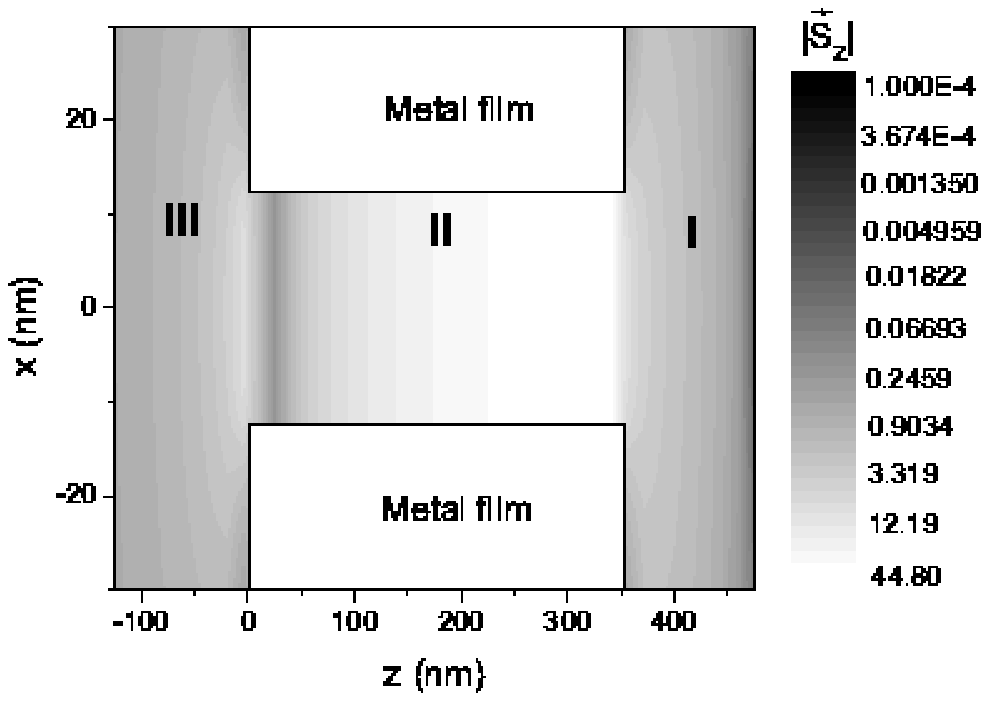}
\subfigure{\label{fig:4a}(a)}
\includegraphics[keepaspectratio,width=\columnwidth]{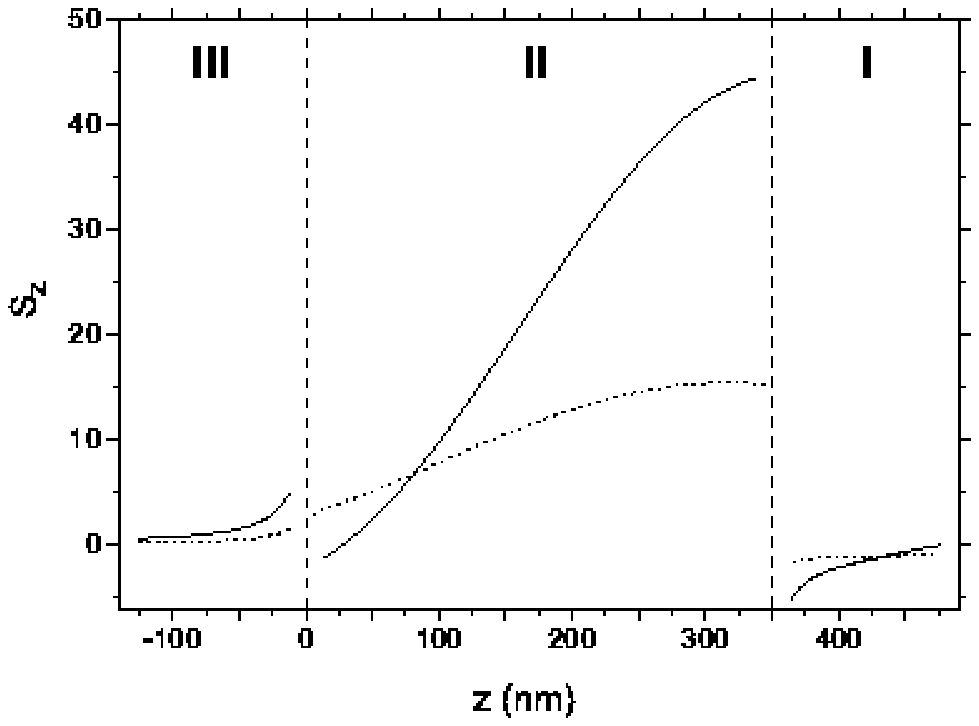}
\subfigure{\label{fig:4b}(b)} \caption{\label{fig:4}(a) The
spatial distribution $|\vec{S}_z(x,z)|$ of the absolute value of
the normalized energy flux along the $z$ direction inside and
outside the slit. The distribution $|S_z(x,z)|$ is shown in the
logarithmical scale. The flux $S_z$ is given by the normalized
fluxes $S^{ird}(U^i,U^r,U^d)/S^i(U^i)$, $S^2(U^2)/S^i(U^i)$ and
$S^3(U^3)/S^i(U^i)$ for the regions I, II and III, respectively.
(b) The energy flux distribution $S_z(x,z)$ at $x=0$. The slit
width $2a=25$~nm, the film thickness $b=351$~nm and the wavelength
$\lambda=800$~nm (solid line) and $\lambda = 882$~nm (dotted line)
corresponding to a transmission resonance and a little reflection
resonance, respectively.}
\end{figure}

The dispersions $S_{int}^d(\lambda)$, $S_{int}^{ird}(\lambda)$ and
$S_{int}^3(\lambda)=T(\lambda)$ shown in Fig.~\ref{fig:2a}
indicate the wave-cavity interaction behavior, which is similar to
that in the case of a Fabry-Perot resonator. The fluxes
$S_{int}^d(U^d)$, $S_{int}^{ird}(U^i,U^r,U^d)$ and
$S_{int}^3(U^3)=T$ exhibit the Fabry-Perot-like maxima around the
resonance wavelengths $\lambda_r^m\approx{2b/m}$. In order to
understand the connection between the Fabry-Perot-like resonances
and the total reflection, we computed the amplitude and phase
distributions of the light wave at the resonant and near-resonant
wavelengths inside and outside the slit cavity (see,
Figs.~\ref{fig:2b}, \ref{fig:3} and \ref{fig:4}). At the resonance
wavelengths, the intra-slit fields possess maximum amplitudes with
Fabry-Perot-like spatial distributions (Fig.~\ref{fig:2b}).
However, in contrast to the Fabry-Perot-like modal distributions,
the resonant configurations are characterized by antinodes of the
electric field at each open aperture of the slit. Such a behavior
is in agreement with the results\cite{Asti,Lind,Xie}. It is
interesting that at the slit entrance, the amplitudes $E_x$ of the
resonant field configuration possesses the Fabry-Perot-like phase
shift on the value of $\pi$ (Fig.~\ref{fig:3}). The integrated
fluxes $S_{int}^{ird}(U^i,U^r,U^d)$, $S_{int}^d(U^d)$ and
$S_{int}^3(U^3)$, at the first resonant wavelength $\lambda_r^1$,
exhibit enhancement by a factor $\lambda/2\pi{a}\approx{10}$ with
respect to the incident wave (Fig.~\ref{fig:2a}). For the
comparison, the normalized resonant fluxes
$S_n^{ird}(U^i,U^r,U^d)$ and $S_n^3(U^3)$ in the near-field zone
($z\approx{-2a}$) are about 5 times greater than the incident wave
(see, Fig.~\ref{fig:4}). It should be stressed that the resonantly
enhanced intra-cavity intensity $S_n^2(U^2)$ is about 10 times
higher than the resonant fluxes $S_n^{ird}(U^i,U^r,U^d)$ and
$S_n^3(U^3)$ localized in the near field zone in front of the slit
and behind the slit, respectively (Fig.~\ref{fig:4}). The
interference of the incident $U^i(x,z)$ wave and the backward
scattered fields $U^r(x,z)$ and $U^d(x,z)$, at the resonant
wavelengths $\lambda_r^m\approx{2b}/m$, produces in the near-field
diffraction zone a strongly localized wave whose normalized flux
$S_n^{ird}(U^i,U^r,U^d)$ is $\lambda/2\pi{a}\approx$10-10$^3$
times greater than the incident wave, but about one order of
magnitude smaller than the resonant intra-cavity intensity.

In our model we considered an incident wave with TM polarization.
According to the theory of waveguides, the vectorial wave
equations for this polarization can be reduced to one scalar
equation describing the magnetic field $H$ of TM modes. The
electric component $E$ of these modes is found using the field $H$
and Maxwell's equations. The TM scalar equation for the component
$H$ is decoupled from the similar scalar equation describing the
field $E$ of TE (transverse electric) modes. Hence, the formalism
works analogously for TE polarization exchanging the $E$ and $H$
fields.

\section{Summary and conclusion}
In the present paper, the backward scattering of TM-polarized
light by a two-side-open subwavelength slit in a metal film has
been analyzed. We predict that the reflection coefficient versus
wavelength possesses a Fabry-Perot-like dependence that is similar
to the anomalous behavior of transmission. The open slit totally
reflects the light at the near-to-resonance wavelengths. The
resonance wavelengths for the total reflection are somewhat
red-shifted with respect to the transmission resonances. The
wavelengths of both the reflection and transmission resonances are
red-shifted with respect to the Fabry-Perot wavelengths. The sharp
resonant maxima of transmission are accompanied by the wide minima
of the reflection. In addition, we showed that the interference of
incident and resonantly backward-scattered light produces in the
near-field diffraction zone a strongly localized wave whose
intensity is greater than the incident wave by a factor
$\lambda/2\pi{a}\approx$10-10$^3$ and about one order of magnitude
smaller than the intra-cavity intensity. The correlation between
the amplitude and phase distributions of light waves inside and
outside the slit was also investigated. The slit cavity was
compared with a Fabry-Perot resonator. We showed that the
amplitude and phase of the resonant wave at the slit entrance and
exit are different from that of a Fabry-Perot cavity. The physical
mechanism responsible for the total reflection is the interference
of the backward diffracted resonant field $U^d(x,z)$ and the
reflected non-resonant field $U^r(x,z)$ in the energy flux at the
near-to-resonance wavelengths (Fano-type effect). The
wavelength-selective total reflection of light by two-side-open
metal slits may find application in many kinds of sensors and
actuators. The (10-10$^3$)-times and ($10^2$-$10^4$)-times
enhancement of the light intensity in front of the slit and inside
the slit can be used in reflective nanooptics and in intra-cavity
spectroscopy of single atoms. We believe that the presented
results gain insight into the physics of resonant scattering of
light by subwavelength nano-slits in metal films.
\begin{acknowledgments}
The authors appreciate the valuable comments and suggestions of
the anonymous referees. This study was supported by the Fifth
Framework of the European Commission (Financial support from the
EC for shared-cost RTD actions: research and technological
development projects, demonstration projects and combined
projects. Contract NG6RD-CT-2001-00602) and in part by the
Hungarian Scientific Research Foundation (OTKA, Contracts T046811
and M045644) and the Hungarian R{\&}D Office (KPI, Contract
GVOP-3.2.1.-2004-04-0166/3.0).
\end{acknowledgments}

\section*{Appendix}

\newcommand{\ead}[1]{\mathrm{e}^{#1}}

We briefly describe the Neerhoff and Mur model\cite{Neer,Betz1} of
the scattering of a plane continuous wave by a subwavelength slit
of width $2a$ in a perfectly conducting metal screen of thickness
$b$. The slit is illuminated by a normally incident plane wave
under TM polarization (magnetic-field vector parallel to the
slit), as shown in Fig.~\ref{fig:5}. The magnetic field of the
wave is assumed to be time harmonic and constant in the $y$
direction:
\begin{eqnarray}
\vec{H}(x,y,z,t)=U(x,z)\ead{-i\omega{t}}\vec{e}_y.
\label{eq:1}
\end{eqnarray}
The electric field of the wave is found from the scalar field $U(x,z)$
using Maxwell's equations.
The restrictions in Eq.~(\ref{eq:1}) reduce the diffraction problem to one involving
a single scalar field $U(x,z)$ in two dimensions. The field is represented by $U_{j}(x,z)$
($j$=1,2,3 in region I, II and III, respectively), and satisfies the
Helmholtz equation: $({\nabla}^2+k_{j}^2)U_j=0,$
where $j=1,2,3$. In region I, the field $U_{1}(x,z)$ is decomposed into three
components:
\begin{equation}
U_1(x,z)=U^i(x,z)+U^r(x,z)+U^d(x,z),
\end{equation}
each of which satisfies the Helmholtz equation. $U^i$ represents the incident
field:
\begin{eqnarray}
U^i(x,z)=\ead{-ik_1z}.
\end{eqnarray}
$U^r$ denotes the reflected field without a slit:
\begin{eqnarray}
U^r(x,z)=U^i(x,2b-z).
\end{eqnarray}
Finally, $U^d$ describes the diffracted field in region I due to the presence of the slit.
With the above set of equations and standard boundary conditions
for a perfectly conducting screen, a unique solution exists for
the scattering problem. The solution is found by using the Green
function formalism.

The magnetic ${\vec{H}}(x,z,t)$ fields in regions I, II, and III
are given by
\begin{eqnarray}
H^1(x,z)=\exp(-ik_1z)+\exp(-ik_1(2b-z))\nonumber\\
\frac{ia}{N}\frac{\epsilon_1}{\epsilon_2}\sum_{j=1}^{N}H^1_0(k_1\sqrt{(x-x_j)^2+(z-b)^2})(DU_b)_j,
\end{eqnarray}
\begin{eqnarray}
\lefteqn{H^2(x,z)=-\frac{i}{2N\sqrt{k_2^2}}\ead{i\sqrt{k_2^2}|z|}\sum_{j=1}^{N}(DU_0)_j+\frac{i}{2N\sqrt{k_2^2}}}\nonumber\\
&&\times\ead{i\sqrt{k_2^2}|z-b|}\sum_{j=1}^N(DU_b)_j-\frac{1}{2N}\ead{i\sqrt{k_2^2}|z|}\sum_{j=1}^{N}(U_0)_j+\frac{1}{2N}\nonumber\\
&&\times\ead{i\sqrt{k_2^2}|z-b|}\sum_{j=1}^{N}(U_b)_j-\frac{i}{N}\sum_{m=1}^{\infty}\frac{1}{\gamma_1}\cos{\frac{m\pi(x+a)}{2a}}\ead{i\gamma_1|z|}\nonumber\\
&&\times\sum_{j=1}^{N}\cos\frac{m\pi(x_j+a)}{2a}(DU_0)_j-\frac{1}{N}\sum_{m=1}^{\infty}\cos\frac{m\pi(x+a)}{2a}\nonumber\\
&&\times\ead{i\gamma_1|z|}\sum_{j=1}^N\cos\frac{m\pi(x_j+a)}{2a}(U_0)_j\nonumber+\frac{i}{N}\sum_{m=1}^{\infty}\frac{1}{\gamma_1}\ead{i\gamma_1|z-b|}\\
&&\times\cos\frac{m\pi(x+a)}{2a}\sum_{j=1}^N\cos\frac{m\pi(x_j+a)}{2a}(DU_b)_j\nonumber\\
&&+\frac{1}{N}\sum_{m=1}^{\infty}\cos(m\pi\frac{x+a}{2a})\ead{i\gamma_1|z-b|}\nonumber\\
&&\times\sum_{j=1}^N\cos\frac{m\pi(x_j+a)}{2a}(U_b)_j,
\end{eqnarray}
\begin{eqnarray}
\lefteqn{H^3(x,z)=i\epsilon_3\sum_{j=1}^N\frac{a}{N\epsilon_2}(D\vec{U}_0)_j}\nonumber\\
&&\times H_0^{(1)}\left[k_3\sqrt{(x-x_j)^2+z^2}\right],
\end{eqnarray}
where $x_{j}=2a(j-1/2)/N-a$, $j=1,2,\dots,N$; $N>2a/z$;
$H_0^{(1)}(X)$ is the Hankel function; $\vec{H}^i=H^i\cdot
\vec{e}_y$,  $i=1,2,3$; $\gamma_m=[k_2^2-(m{\pi}/2a)^2]^{1/2}$.
The coefficients $(D{\vec{U}}_0)_{j}$ are found by solving
numerically four coupled integral equations.
For more details on the model and the numerical solution of the
Neerhoff and Mur coupled integral equations,
see the references\cite{Neer,Betz1}.


\begin{references}
\bibitem{Neer}F. L. Neerhoff and G. Mur, Appl.\ Sci.\ Res.\ {\bf28}, 73 (1973).
\bibitem{Harr}R.F. Harrington and D.T. Auckland, IEEE\ Trans.\ Antennas\ Propag\ {\bf{AP28}},
616 (1980).
\bibitem{Betz1}E. Betzig, A. Harootunian, A. Lewis, and M. Isaacson, Appl.\ Opt.\ {\bf25},
1890 (1986).
\bibitem{Ebbe}T.W. Ebbesen, H.J. Lezec, H.F. Ghaemi, T. Thio, and P.A. Wolff, Nature\
(London)\ {\bf391}, 667 (1998).
\bibitem{Hess}A. Hessel and A.A. Oliner, Appl.\ Opt.\ {\bf4}, 1275 (1965).
\bibitem{Nev}M. Nevi\`ere, D. Maystre, and P. Vincent, J.\ Opt.\ {\bf8}, 231 (1977).
\bibitem{Nev1}D. Maystre and M. Nevi\`ere, J.\ Opt.\ {\bf8}, 165 (1977).
\bibitem{Sarr1}M. Sarrazin, J.P. Vigneron, and J.M. Vigoureux, Phys.\  Rev.\ B\ {\bf67}, 085415 (2003).
\bibitem{Port}J. A. Porto, F.J. Garc\'ia-Vidal, and J.B. Pendry, Phys.\ Rev.\ Lett.\
{\bf83}, 2845 (1999).
\bibitem{Asti}S. Astilean, Ph. Lalanne, and M. Palamaru, Opt.\ Commun.\
{\bf175}, 265 (2000).
\bibitem{Hibb}A.P. Hibbins, J.R. Sambles and C.R. Lawrence, Appl.\ Phys.\ Lett.\ {\bf81},
4661 (2002).
\bibitem{Cao}Q. Cao and P. Lalanne, Phys.\ Rev.\ Lett.\ {\bf88}, 057403 (2002).
\bibitem{Taka}Y. Takakura, Phys.\ Rev.\ Lett.\ {\bf86}, 5601 (2001).
\bibitem{Yang}F.Z. Yang and J.R. Sambles, Phys.\ Rev.\ Lett.\ {\bf89}, 063901 (2002).
\bibitem{Kuk2}S.V. Kukhlevsky, M. Mechler, L. Csapo, K. Janssens, and O. Samek, Phys.\ Rev.\ B\ {\bf70}, 195428 (2004).
\bibitem{Gar1}F.J. Garc\'ia-Vidal, H.J. Lezec, T.W. Ebbesen, and L. Mart\'in-Moreno, Phys.\ Rev.\ Lett.\ {\bf90}, 231901
(2003).
\bibitem{Bori}A.G. Borisov, F.G. Garcia de Abajo, and S.V. Shabanov, Phys.\ Rev.\ B\ {\bf71}, 075408 (2005).
\bibitem{Stee}J.M. Steele, C.E. Moran, A. Lee, C.M. Aguirre, and N.J. Halas, Phys.\ Rev.\ B\
{\bf68}, 205103 (2003).
\bibitem{Gar3}F.J. Garc\'ia-Vidal and L. Mart\'in-Moreno, Phys.\ Rev.\ B\ {\bf66}, 155412 (2002).
\bibitem{Lind}J. Lindberg, K. Lindfors, T. Setala, M. Kaivola, and A.T. Friberg, Opt.\ Express\ {\bf12}, 623
(2004).
\bibitem{Xie}Y. Xie, A.R. Zakharian, J.V. Moloney, and M. Mansuripur, Opt.\ Express\ {\bf12}, 6106 (2004).
\bibitem{Cscher}U. Schr\"oter and D. Heitmann, Phys.\ Rev.\ B\ {\bf58}, 15419 (1998).
\bibitem{Trea}M.M.J. Treacy, Phys.\ Rev.\ Lett.\ {\bf75}, 606 (1999).
\bibitem{Vigo} J.M. Vigoureux and R. Giust, Opt.\ Commun.\ {\bf186}, 21 (2000).
\bibitem{Monz} J.J. Monz\'on, T. Yonte, and  L.L. S\'anchez-Soto, Opt.\ Commun.\ {\bf218}, 43 (2003)
\bibitem{Pop}E. Popov, M. Nevi\`ere, S. Enoch, and R.
Reinisch, Phys.\ Rev.\ B\ {\bf62}, 16100 (2000).
\bibitem{Bozs}S.I. Bozhevolnyi, J. Erland, K. Leosson, P.M.W. Skovgaard, and J.M.
Hvam, Phys.\ Rev.\ Lett.\ {\bf86}, 3008 (2001).
\bibitem{Barb}A. Barbara, P. Quemerais, E. Bustarret, and T. Lopez-Rios, Phys.\ Rev.\ B\
{\bf66}, 161403 (2002).
\bibitem{Dykh}A.M. Dykhne, A.K. Sarychev, and V.M. Shalaev, Phys.\ Rev.\ B\ {\bf67},
195402 (2003).
\bibitem{Shi}X.L. Shi, L. Hesselink, and R.L. Thornton, Opt.\ Lett.\ {\bf28}, 1320 (2003).
\bibitem{Scho}H.F. Schouten, T.D. Visser, D. Lenstra, and H. Blok, Phys.\ Rev.\ E\ {\bf67}, 036608 (2003).
\bibitem{Naha}A. Nahata, R.A. Linke, T. Ishi, and K. Ohashi, Opt.\ Lett.\ {\bf28}, 423 (2003).
\bibitem{Bouh}A. Bouhelier, M. Beversluis, A. Hartschuh, and L. Novotny,
Phys.\ Rev.\ Lett.\ {\bf90}, 013903 (2003).
\bibitem{Li}K.R. Li, M.I. Stockman, and D.J. Bergman, Phys.\  Rev.\ Lett.\ {\bf91}, 227402 (2003).
\bibitem{Decha}A. Dechant and A.Y. Elezzabi, Appl.\ Phys.\ Lett.\ {\bf84}, 4678 (2004).
\bibitem{Fan}W.J. Fan, S. Zhang, B. Minhas, K.J. Malloy, and S.R.J. Brueck, Phys.\ Rev.\ Lett.\ {\bf94}, 033902 (2005).
\bibitem{Zay}A.V. Zayats, I.I. Smolyaninov, and A.A Maradudin, Phys.\  Rep.\ {\bf408}, 131 (2005).
\bibitem{Lab}M. Labardi, M. Zavelani-Rossi, D. Polli, G. Cerullo, M. Allegrini, S. De Silvestri,
and O. Svelto, Appl.\ Phys.\  Lett.\ {\bf86}, 031105 (2005).
\bibitem{Ben}Y. Ben-Aryeh, International\ J.\ Quantum\ Information\ {\bf3}, 111 (2005).


\end{references}
\end{document}